\documentclass[aps,showpacs,eqsecnum]{revtex4}

\usepackage{epsfig}
\usepackage{graphicx}
\usepackage{amsmath}
\usepackage{amssymb}
\usepackage{color}

\newif\ifold             \oldtrue            
\def\ba{\begin{eqnarray}}
\def\ea{\end{eqnarray}}

\newcommand{\be}{\begin{equation}}
\newcommand{\ee}{\end{equation}}

\begin{document}

\title{Supercritical electric dipole and migration of electron wave function in gapped graphene}
\date{\today}

\author{E. V. Gorbar}
\affiliation{Department of Physics, Taras Shevchenko National University of Kiev, Kiev, 03680, Ukraine}
\affiliation{Bogolyubov Institute for Theoretical Physics, Kiev, 03680, Ukraine}

\author{V. P. Gusynin}
\affiliation{Bogolyubov Institute for Theoretical Physics, Kiev, 03680, Ukraine}

\author{O. O. Sobol}
\affiliation{Department of Physics, Taras Shevchenko National University of Kiev, Kiev, 03680, Ukraine}

\begin{abstract}
We study the Dirac equation for quasiparticles in gapped graphene with two oppositely charged impurities
by using the technique of linear combination of atomic orbitals and variational Galerkin--Kantorovich method. We show that for sufficiently large charges of impurities the wave function of the occupied 
electron bound state of the highest energy  changes its localization from the negatively charged impurity 
to the positively charged one as the distance between the impurities increases. This migration of the electron wave function of supercritical electric dipole is a generalization of the familiar phenomenon of the atomic collapse of single charged impurity to the case where electron-hole pairs are spontaneously created from vacuum in  bound states with charge impurities thus partially screening them.
\end{abstract}
\maketitle

\section{Introduction} 
Atomic collapse or supercritical Coulomb center instability is a well-known phenomenon in quantum electrodynamics. For the electron in the Coulomb field of a nucleus of charge $Ze$, atomic collapse takes place for $Z \gtrsim 170$ \cite{Zeldovich,Greiner} when the lowest energy electron bound state dives into the lower continuum. This leads to the spontaneous creation of electron-positron pairs with the electrons screening the positively charged nucleus and the positrons emitted to infinity. Since supercritically charged nuclei are not encountered in nature, this phenomenon was never observed in QED. The situation essentially changes in graphene whose effective coupling constant $e^2/(\hbar v_F)\approx 2.2$ 
($v_F \approx c/300$ is the Fermi velocity) exceeds unity that drastically decreases the value of the critical charge in graphene \cite{Shytov,Pereira,Novikov}.

Experimental observation of the supercritical instability  for charge impurities in graphene  remained elusive until recently. One can reach the supercritical regime by collecting a large enough number of charged adatoms in a certain region of graphene. Such an approach was recently successfully realized \cite{Wang} by using a STM tip  in order to create clusters of charged calcium dimers.

The study of supercritical instability of one Coulomb center in the continuum model in gapped graphene
was extended  \cite{two-centers} to the case of the simplest cluster of two equally charged impurities
when the charges of impurities are subcritical, whereas their total charge exceeds a critical one. We determined the critical distance between the impurities separating the supercritical and subcritical
regimes as a function of the charges of impurities and the gap.

An interesting problem of two oppositely charged impurities in graphene was recently considered in Refs.\cite{Egger,Matrasulov}. By studying the Dirac equation for quasiparticles in graphene in
the field of point electric dipole, it was shown that this equation admits towers of infinitely many
bound states exhibiting a universal Efimov-like scaling hierarchy. The electric dipole problem is 
obviously particle-hole symmetric one and, consequently, the energy levels are symmetric with respect 
to the change of the sign of energy $E \to - E$. Therefore, naively the bound electron and hole states 
can cross as the dipole moment increases only at $E=0$. However, such a level crossing is impossible 
for two states with the same quantum numbers in view of the avoided crossing theorem \cite{Wigner}. 
Indeed, the explicit calculations in Ref.\cite{Egger} showed that the positive and negative energy 
levels first approach each other and then go away. Since no electron bound state dives into the lower
continuum as the electric dipole moment increases, the conclusion was made in Ref.\cite{Egger} that supercriticality is unlikely to occur in the electric dipole problem.

In this paper, we address the problem of supercritical instability for quasiparticles in graphene with
two oppositely charged impurities situated at finite distance $R$ ({\it finite} electric dipole). We
show that a new type of instability is revealed in this problem connected with the change of localization (migration) of the electron wave function on impurities as the distance between the two impurities changes. This migration leads to the spontaneous creation of an electron-hole pair with the electron and hole partially screening the positively and negatively charged impurities, respectively. The necessary condition for the instability to occur is the crossing of the energy levels of the corresponding single positively
and negatively charged impurities.

\section{Dirac equation} 
The electron quasiparticle states in the vicinity of the $K_{\pm}$ points of graphene in the field of two oppositely charged impurities are described by the following Dirac Hamiltonian in $2+1$ dimensions (we set $\hbar=1$ in what follows):
\begin{equation}
H=v_{F}\boldsymbol{\sigma}\boldsymbol{p}+\xi\Delta\sigma_z-eV(\mathbf{r}),
\label{general-Hamiltonian}
\end{equation}
where $-e<0$ is the electron charge, $\boldsymbol{p}$ is the two-dimensional canonical momentum, $\sigma_{i}$ are the Pauli matrices, and $\Delta$ is a quasiparticle gap. The gap in graphene can be
opened in various ways, e.g., by depositing graphene on a substrate \cite{Ponomarenko,Song} or simply
due to finite-size effects in graphene nanoribbons \cite{Peres}. Hamiltonian (\ref{general-Hamiltonian}) acts on two-component spinor $\Psi_{\xi s}$ which carries the valley ($\xi=\pm)$ and spin ($s=\pm$) indices. We will use the standard convention: $\Psi^{T}_{+s}=(\psi_{A},\psi_{B})_{K_{+}s}$, whereas $\Psi^{T}_{-s}=(\psi_{B},\psi_{A})_{K_{-}s}$, and $A,B$ refer to two sublattices of hexagonal graphene lattice. The regularized interaction dipole potential for charged impurities $\pm Q$, $Q=Ze$, situated at $(\pm R/2,0$) in the $(x,y)$ plane is given by
\begin{equation}
V\left(\mathbf{r}\right)=\frac{Q}{\kappa}\hspace{-1mm}\left(\hspace{-1mm}\frac{1}{\sqrt{(x+R/2)^{2}+y^{2}
+r_{0}^{2}}}-\left(R \rightarrow -R\right)\hspace{-1mm}\right),
\label{dipole-potential}
\end{equation}
where $\kappa$ is the dielectric constant and the Coulomb potential of each impurity is regularized by $r_0$, which is of the order of the graphene lattice spacing. Since the interaction potential (\ref{dipole-potential}) does not depend on spin, we will omit the spin index $s$ in wave functions in
what follows. Further, for the sake of definiteness, we will consider electrons only in the $K_+$ valley (the Dirac equation for the electrons in the $K_-$ valley is obtained by replacing $\Delta$ with $-\Delta$). The main difficulty in solving the Dirac equation for the electron in the electric dipole potential is
that variables in this problem are not separable in any known orthogonal coordinate system. Therefore,
we will utilize in this paper the technique of linear combination of atomic orbitals (LCAO) and variational Galerkin--Kantorovich (GK) method.

Hamiltonian (\ref{general-Hamiltonian})  has a particle-hole symmetry expressed by $\Omega H\Omega^+=-H$, where the unitary operator $\Omega=\sigma_x {\cal R}_x$ with the operator ${\cal R}_x$ of reflection $x \rightarrow -x$ satisfies $\Omega^2=1$. It follows then that an eigenstate $\Psi_E(x,y)$ with energy $E$
has a partner $\Psi_{-E}(x,y)=\Omega\Psi_E(x,y)=\sigma_x\Psi_E(-x,y)$ with energy $-E$, hence, all
solutions to the Dirac equation come in pairs with $\pm E$.

The Dirac Hamiltonian with the electric dipole potential commutes with the operator
$U=\sigma_{z}K\mathcal{R}_{y}$ where $K$ is the complex conjugation, $\mathcal{R}_{y}$ is the operator
of reflection $y \rightarrow -y$. Since $U^{2}=1$, wave functions are split into two classes
$U|\Psi_{\lambda}\rangle=\lambda|\Psi_{\lambda}\rangle$, where $\lambda=\pm 1$. Since the operator $U$ 
is antilinear, the function $|\Psi_{-}\rangle$ related to the function $|\Psi_{+}\rangle$ by means of the phase transformation, $|\Psi_{-}\rangle=i|\Psi_{+}\rangle$, is an eigenfunction with $\lambda=-1$. Therefore, there is no need to consider functions $|\Psi_{-}\rangle$. Hence  the components of wave
functions $|\Psi_{+}\rangle =(\phi,\chi)^T$, satisfy the constraint conditions $\phi^{*}(-y)=\phi(y),\chi^{*}(-y)=-\chi(y)$ which are consistent
with the Dirac equation.

It is convenient to work with dimensionless quantities $h=\frac{H}{\Delta}$ and $\epsilon=\frac{E}{\Delta}$ and use dimensionless coordinates and distances defined in units of $R_{\Delta}=\frac{\hbar v_{F}}{\Delta}$. We introduce also dimensionless coupling constant $\zeta=\frac{eQ}{\hbar v_{F}\kappa}$. Thus, Dirac equation reduces to a system of two coupled ordinary differential equations of the first order
\begin{equation}
\left\{
\begin{array}{l}
-i(\partial_{x}+i\partial_{y})\phi+(v-\epsilon-1)\chi=0,\\
-i(\partial_{x}-i\partial_{y})\chi+(v-\epsilon+1)\phi=0,
\end{array}
\right.
\label{system-equations}
\end{equation}
where $v(r)=-eV(r)/\Delta$.
We will find numerical solutions of these equations by using the GK method in the class of wave functions with $\lambda=1$. Still it is instructive to analyse the Dirac equation analytically. For this, we utilize the LCAO method.

\section{LCAO technique} The LCAO method is well known and widely used in molecular physics \cite{Diu}.
Wave functions in this method are chosen as linear combinations of basis functions, where the latter
are usually the electron functions centered on the corresponding atoms of the molecule. By minimizing
the total energy of the system, the coefficients of the linear combinations are then determined. The
LCAO method was recently used in Ref.\cite{Matrasulov} to solve the two Coulomb centers problem in
graphene and the corresponding results are similar to those found by matching of the asymptotics \cite{two-centers}. However, the LCAO method was not applied to the analysis of the electric dipole problem performed in Refs.\cite{Egger,Matrasulov}.

As to the atomic orbitals for problem (\ref{system-equations}), we take the wave function of the lowest energy bound state in the field of positively charged impurity and the wave function of the highest energy bound state for negatively charged impurity (these wave  functions are related to each other by charge conjugation).

We begin our analysis with the Dirac equation for the electron in graphene with one positively
charged impurity $h_{p}\Psi_p=\epsilon\Psi_p$ with Hamiltonian
\begin{equation}
h_p=-i(\sigma_{x}\partial_{x}+\sigma_{y}\partial_{y})+\sigma_{z}-\frac{\zeta}{\sqrt{r^{2}+r_{0}^{2}}}.
\end{equation}
[The Hamiltonian $h_n$ for the electron in the field of negatively charged impurity is obtained
from the Hamiltonian $h_p$ by the change of the sign of the last term in $h_p$.]

We determine numerically the energy levels by using the shooting method with regular boundary conditions
at $r=0$ for the wave functions and requiring that the wave functions decrease at infinity.
\begin{figure}[ht]
  \centering
  \includegraphics[scale=0.32]{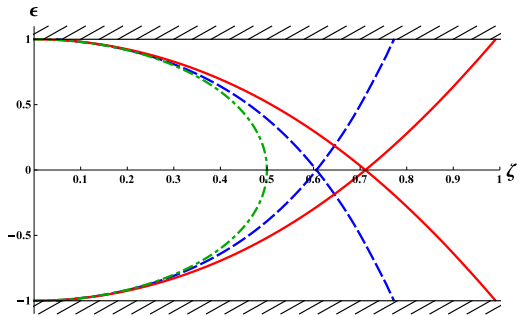}
  \caption{(Color online) The energy of the electron bound state with $j=1/2$ in the regularized
  Coulomb potential with $\pm Q$ charges  as a function of $\zeta$ for different values of the regularization parameter: $r_{0}=0.05 R_{\Delta}$
  (red solid lines), $r_{0}=0.01 R_{\Delta}$ (blue
  dashed lines), $r_{0}=0$ (green dash-dotted lines).}
  \label{fig1-LCAO}
\end{figure}
The energy of the lowest (highest) electron bound state with the total angular momentum $j=1/2$ in the regularized Coulomb potential with the charge $+Q$ ($-Q$) is plotted in Fig.\ref{fig1-LCAO} for different values of the regularization parameter $r_{0}$ as a function of $\zeta$. The levels which descend from
the upper continuum correspond to the positive charge $+Q$ while those which are pushed from the lower continuum and grow with $\zeta$ correspond to the negative charge $-Q$. These results are in accordance 
with calculations in Ref.\cite{excitonic-instability} (see Fig.4 there) where slightly different regularization for the one Coulomb center potential was used which admitted an analytical solution. They also reproduce qualitatively the behavior seen directly at the tight-binding level on a honeycomb lattice \cite{Kotov}. For nonregularized Coulomb potential with positive charge the lowest bound-state energy is always positive, it reaches the value $\epsilon=0$ for $\zeta=1/2$ and becomes purely imaginary for $\zeta>1/2$ (the fall into the center phenomenon \cite{Shytov,Pereira,Novikov}). For regularized Coulomb potential with charge $+Q (-Q)$, the lowest (highest) bound-state energy crosses $\epsilon=0$ and  dives into the lower (upper) continuum at certain value of the charge. For example, for $r_{0}=0.05 R_{\Delta}$, this happens for $\zeta\approx1$.

Since the operator $U_{c}=\sigma_x K$ interchanges the $h_p$ and $h_n$ Hamiltonians, $U_{c}h_pU^{+}_{c}=-h_n$, the electron levels in the field of negatively charged center described by the Hamiltonian $h_n$ are obtained by the reflection $\epsilon\rightarrow -\epsilon$ and intersect
with the levels of the Hamiltonian $h_p$ at $\epsilon=0$. The corresponding critical value $\zeta_c$ when this happens will play a crucial role in the behavior of energy levels in the dipole potential because the behavior of these levels dramatically changes depending on whether $\zeta<\zeta_c$ or $\zeta>\zeta_c$.
For chosen values of the regularization parameter $r_0$ in Fig.\ref{fig1-LCAO}, the critical coupling $\zeta_c=0.6\, (r_0=0.01R_\Delta)$ and $\zeta_c=0.7\, (r_0=0.05R_\Delta)$. In general,  $\zeta_c$
increases with the increase of $r_0$.

\begin{figure}[ht]
  \centering
  \includegraphics[scale=0.4]{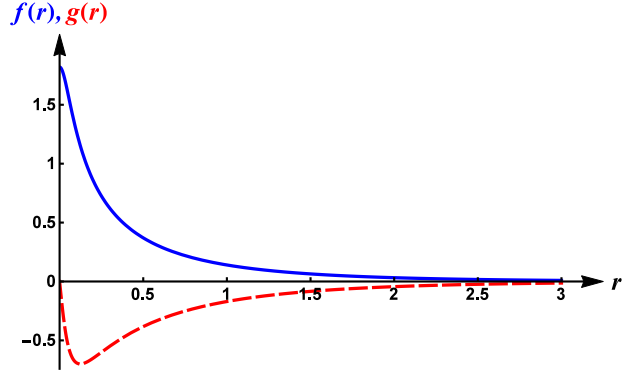}
  \caption{(Color online) The radial functions in the regularized potential of one positively charged
  impurity with $\zeta=0.85$, $r_{0}=0.05 R_{\Delta}$: the function $f(r)$ is shown by blue solid line and $g(r)$ by red dashed line.}
  \label{rad-fcs}
  \end{figure}

We are ready now to consider the Dirac equation for quasiparticles in graphene with two oppositely charged impurities. The corresponding Hamiltonian has the form
\begin{equation}
h=-i(\sigma_{x}\partial_{x}+\sigma_{y}\partial_{y})+\sigma_{z}+
\frac{\zeta}{\sqrt{r_{n}^{2}+r_{0}^{2}}}-\frac{\zeta}{\sqrt{r_{p}^{2}+r_{0}^{2}}},
\end{equation}
where $r_{p,n}=\sqrt{(x\pm R/2)^{2}+y^2}$. We seek the wave function as a linear combination (hybridization),
\begin{equation}
\label{linear_combination}
|\Psi\rangle=v_{p}|\Psi_p\rangle+v_{n}|\Psi_n\rangle,
\end{equation}
of the wave functions $\Psi_p$ and $\Psi_n$ which are eigenstates of the Hamiltonians $h_p$ and $h_n$, respectively, with eigenvalues $\pm\epsilon_{0}$, where $\epsilon_{0}$ is the energy of the lowest energy electron bound state in the field of one Coulomb center with the charge $+Q$. Explicitly, the functions $|\Psi_p\rangle, |\Psi_n\rangle$ with the total angular momentum $j=1/2$ are given in polar coordinates
by
\begin{equation}
\Psi_p=\left(\begin{array}{c}f(r_p)\\-i e^{i\theta_p}g(r_p)\end{array}\right),\quad\Psi_n=\left(\begin{array}{c}e^{-i\theta_n}g(r_n)\\
-i f(r_n)\end{array}\right),
\end{equation}
where $\exp[-i\theta_{p,n}]=(x\pm R/2-i y)/r_{p,n}$. The radial functions $f(r), g(r)$ are computed
numerically in the regularized potential of one positively charged impurity (see Fig.\ref{rad-fcs}).

It is substantial for our analysis below to use the electron wave functions $|\Psi_p\rangle, |\Psi_n\rangle$ in the field of single regularized Coulomb centers whose energies may cross zero. By making use of the wave function (\ref{linear_combination}), we project the Dirac equation $h|\Psi\rangle=\epsilon|\Psi\rangle$ on the states $|\Psi_p\rangle$ and $|\Psi_n\rangle$ and find the following secular equation:
\begin{equation}
\det \left|
\begin{array}{cc}
h_{pp}-\epsilon & h_{pn}-S\epsilon\\
h_{np}-S\epsilon & h_{nn}-\epsilon
\end{array}
\right|=0,
\end{equation}
where $h_{ij}=\langle i|h|j\rangle$, $\langle i|j\rangle=\delta_{ij}$, $i,j=\Psi_p,\Psi_n$,
$S=\langle \Psi_p|\Psi_n\rangle=\langle \Psi_n|\Psi_p\rangle$, $\langle \Psi|\Psi\rangle=1$.
%It is easy to show that the overlap integral $S$ vanishes,
It is easy to see that the overlap integral,
\begin{eqnarray}
S=\int dxdy \left(f(r_p)g(r_n)\frac{x-R/2}{r_n}+f(r_n)g(r_p)\frac{x+R/2}{r_p}\right), \nonumber
\end{eqnarray}
vanishes after changing $x \to -x$ in the first term in the last equality.
\begin{figure}[ht]
  \centering
  \includegraphics[scale=0.32]{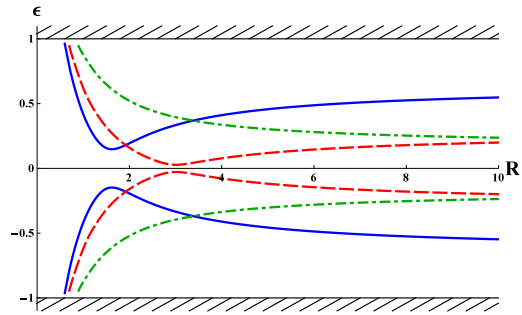}
  \caption{(Color online) The energy of the bound state levels for $r_0=0.05R_\Delta$ as a function of
  a distance in the LCAO method: $\zeta=0.65$ (green dash-dotted lines) and $\zeta=0.8$
(red dashed lines), $\zeta=0.9$ (blue solid lines).}
  \label{fig2-LCAO}
\end{figure}
In order to calculate the coefficients $h_{ij}$, it is convenient to represent the Hamiltonian in the form
$h=h_p+{\zeta}/{\sqrt{r_{n}^{2}+r_{0}^{2}}}=h_n-{\zeta}/{\sqrt{r_{p}^{2}+r_{0}^{2}}}$. Then we find
\begin{eqnarray}
h_{pp}&=&\epsilon_{0}+\langle \Psi_p|\frac{\zeta}{\sqrt{r_{n}^{2}+r_{0}^{2}}}|\Psi_p\rangle
=\epsilon_{0}+\zeta C=-h_{nn},
\label{h-11}\\
h_{pn}&=&-\langle \Psi_p|\frac{\zeta}{\sqrt{r_{p}^{2}+r_{0}^{2}}}|\Psi_n\rangle=-\zeta A=h_{np}.
\label{h-12}
\end{eqnarray}
We find that the Coulomb integral $C$ equals
\begin{eqnarray}
C=\langle p|\frac{1}{\sqrt{r_{n}^{2}+r_{0}^{2}}}|p\rangle=
\int\limits_{0}^{\infty}\frac{4 r dr}{\sqrt{(r+R)^{2}+r_{0}^{2}}} 
{\rm K}\left(\sqrt{\frac{4r R}{(r+R)^{2}+r_{0}^{2}}}\right)\left(f^{2}(r)+g^{2}(r)\right),
\end{eqnarray}
where $K(k)$ is the complete elliptic integral of the first kind. We note that the Coulomb integral
$C$ is positive definite and monotonously decreases with increasing $R$. Further, the resonance
integral $A$ equals
\begin{eqnarray}
A=\langle p|\frac{1}{\sqrt{r_{p}^{2}+r_{0}^{2}}}|n\rangle=
2\int\limits_{-\infty}^{\infty}dx\int\limits_{0}^{\infty}dy \,f(r_{p})g(r_{n}) 
\frac{x- R/2}{r_{n}}\left(\frac{1}{\sqrt{r_{p}^{2}+r_{0}^{2}}}
-\frac{1}{\sqrt{r_{n}^{2}+r_{0}^{2}}}\right).
\end{eqnarray}
The Coulomb integral $C$ and resonance integral $A$  can be computed numerically with the functions $f$
and $g$ found for an isolated impurity problem. Asymptotically at large $R$ they behave as $C\simeq 1/R$
and $A\sim\exp(-\sqrt{1-\epsilon_0^2} R)$.

Finally, we obtain the energy levels
\begin{equation}
\label{spectrum}
\epsilon=\pm\sqrt{(h_{pp})^{2}+(h_{pn})^{2}}=\pm\sqrt{(\epsilon_{0}+\zeta C)^{2}+\zeta^{2}A^{2}},
\end{equation}
which are obviously symmetric with respect to the replacement $\epsilon \to -\epsilon$ in view of
the charge conjugation symmetry of the problem under consideration. We note that the energy levels never cross in agreement with the avoided crossing theorem \cite{Wigner}. Since the Coulomb and resonance integrals, $C$ and $A$, tend to zero as $R\rightarrow \infty$, the energy of the
system for large distances between the impurities tends to $\epsilon\rightarrow\pm |\epsilon_{0}|$ as expected. The coefficients of the wave function of the negative energy level are given by
\begin{eqnarray}
v_p&=&-\frac{h_{pn}}{\sqrt{(h_{np})^{2}+(h_{pp}+\sqrt{(h_{pp})^{2}+(h_{pn})^{2}})^{2}}},\\
v_n&=&\frac{h_{pp}+\sqrt{(h_{pp})^{2}+(h_{pn})^{2}}}{\sqrt{(h_{np})^{2}+(h_{pp}+\sqrt{(h_{pp})^{2}
+(h_{pn})^{2}})^{2}}}.
\end{eqnarray}

\begin{figure}[ht]
  \centering
  \includegraphics[scale=0.16]{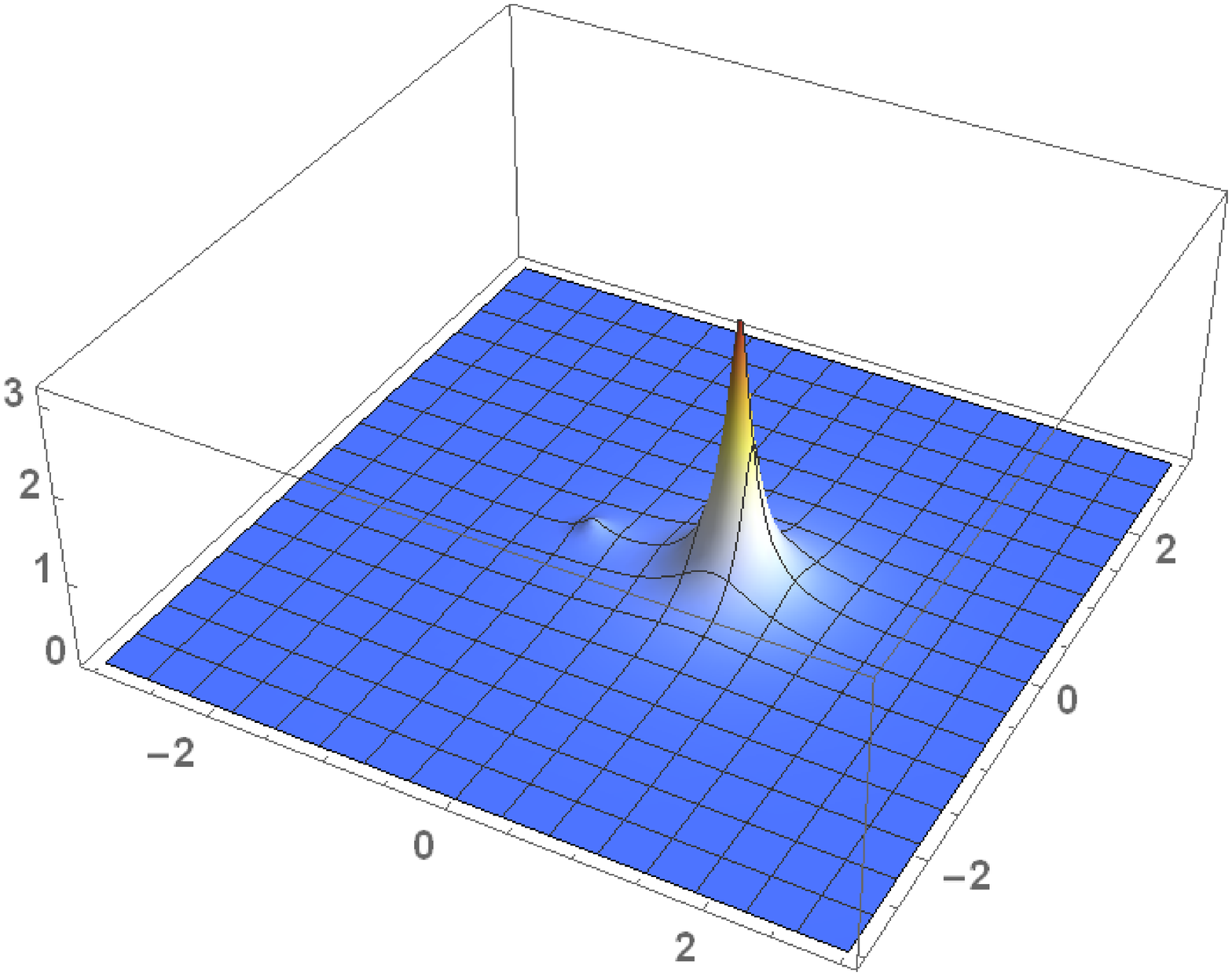}
  \includegraphics[scale=0.16]{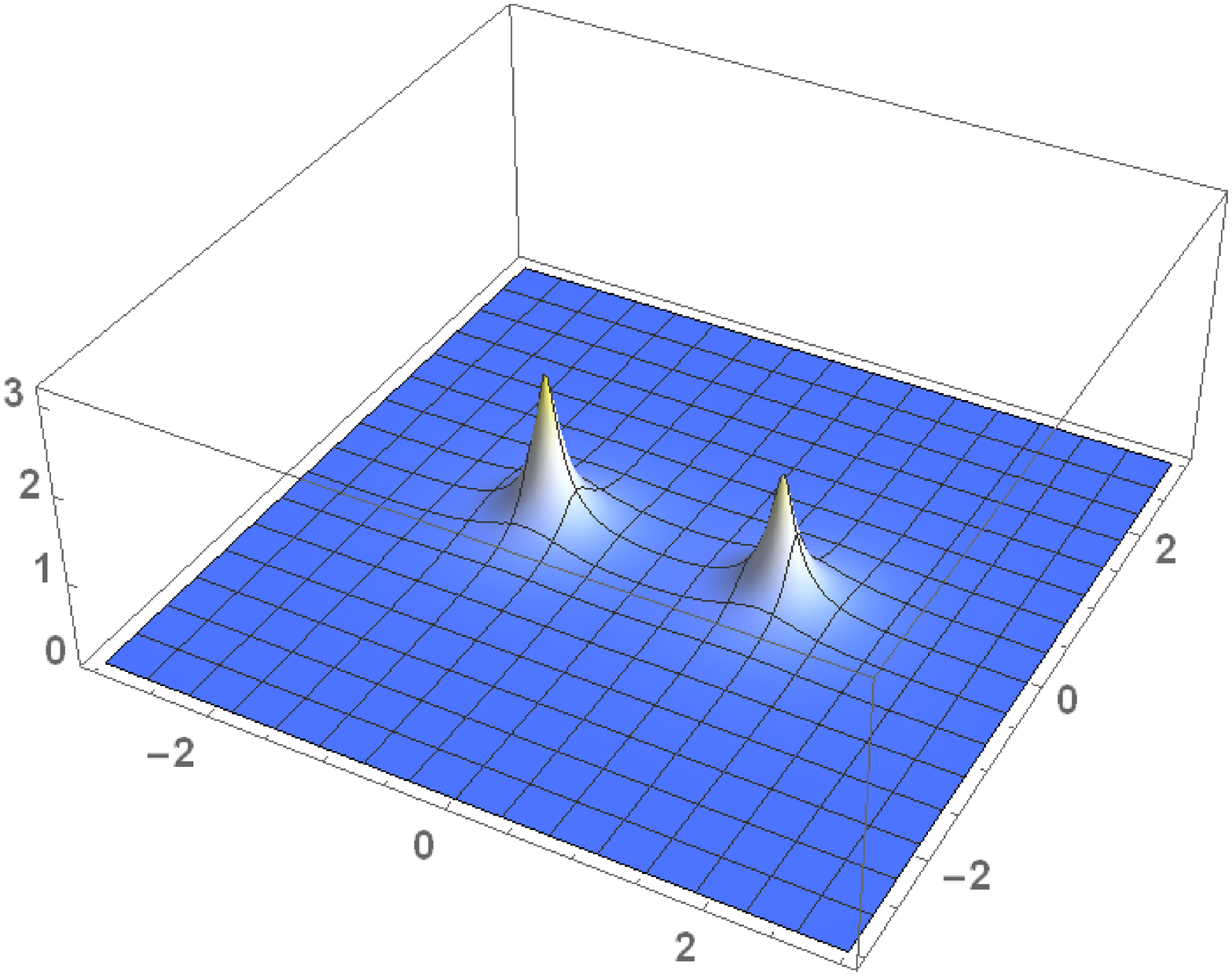}
  \includegraphics[scale=0.16]{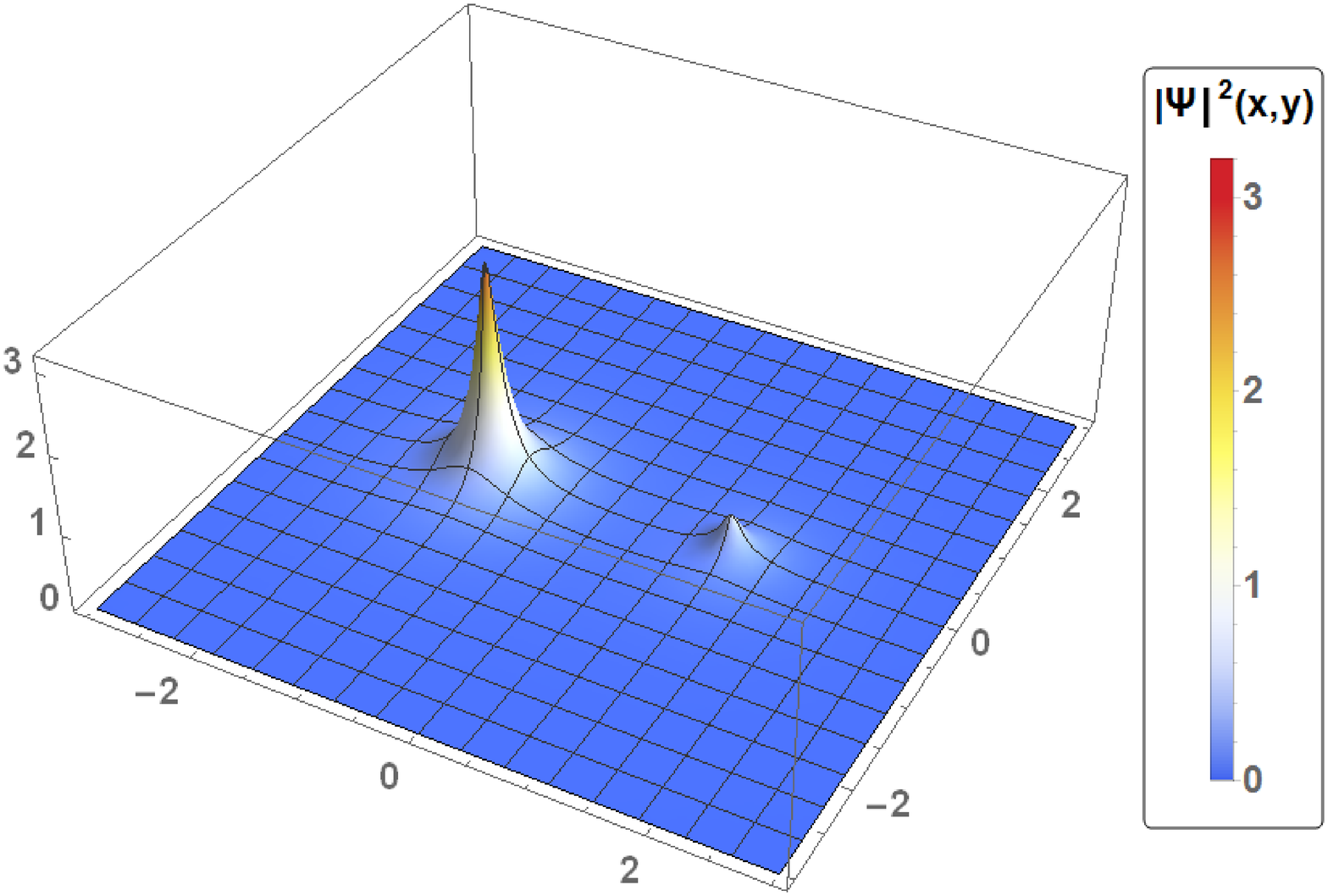}
  \caption{(Color online) The square modulus of the negative energy wave function for $\zeta=0.85$ and
  various distances between the impurities: $R=1.25$ (upper left panel), $R=2.0$ (upper right panel),
  and $R=2.25$ (lower panel) in the LCAO method.}
  \label{wf_LCAO}
\end{figure}
We plot the energy levels of the system for $r_{0}=0.05 R_{\Delta}$ in Fig.\ref{fig2-LCAO} as functions
of $R$ for $\zeta=0.65$, $\zeta=0.8$ and $\zeta=0.9$. In the first case, we have $\zeta<\zeta_c=0.7$ and 
the bound state levels monotonously converge to each other as $R$ increases and never cross. For the  couplings $\zeta=0.8$ and $\zeta=0.9$ which are larger than $\zeta_c=0.7$, their behavior is no longer monotonous. For small $R$, the levels converge like in the previous case. However, after the maximal convergence of the levels they go away with the subsequent increase of $R$. This behavior is typical for 
the avoided crossing \cite{Wigner}.

Eq.(\ref{spectrum}) and the facts that $C$ and $A$ monotonously depend on $R$ and $C\geq 0$ imply that 
the level repulsion can take place only for $\epsilon_{0}<0$ and the energy levels converge most closely 
for $\zeta C=|\epsilon_{0}|$ from which we can determine the corresponding distance $R_{m}$. Exactly at 
this distance $R_{m}$ we have $v_p=v_n=1/\sqrt{2}$ and, consequently, the probability to find the electron near the positively and negatively charged impurities is the same. Furthermore, for $\zeta C \approx |\epsilon_{0}|$, the difference of  the coefficients $v_p$ and $v_n$ squared equals
\begin{eqnarray}
v_n^{2}-v_p^{2}=\frac{2h_{pp}\left(h_{pp}+\sqrt{(h_{pp})^{2}+(h_{pn})^{2}}\right)}
{(h_{np})^{2}+(h_{pp}+\sqrt{(h_{pp})^{2}+(h_{pn})^{2}})^{2}} 
\sim {\rm sign}(h_{pp})={\rm sign}(\epsilon_0 +\zeta C).
\label{difference}
\end{eqnarray}

\begin{figure}[ht]
  \centering
  \includegraphics[scale=0.35]{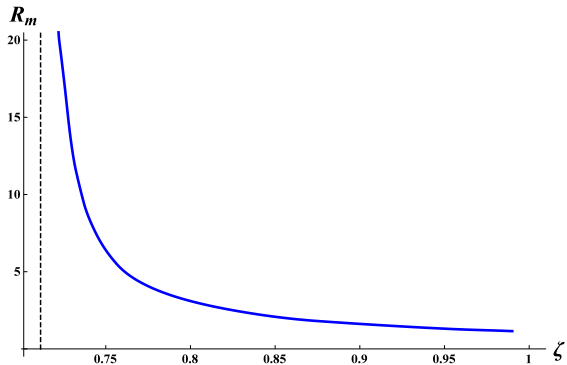}
\caption{The dependence of the distance $R_m$ on the charge $\zeta$ at $r_{0}=0.05 R_{\Delta}$.}
\label{crit_distance}
\end{figure}

The following qualitative picture appears for $\epsilon_0 < 0$. For small $R$, $|v_n| \gg |v_p|$ and, therefore, the electron wave function of the negative energy level in the LCAO method is localized mainly 
on the negatively charged impurity. Although this result seems to be counterintuitive, it is quite natural. The point is that the energy spectrum of the system for $R=0$ is composed of the upper and lower continua. Since the chemical potential in neutral graphene is zero, the electron states of the lower continuum are occupied. For small $R$, the positively charged impurity produces electron bound states which descend from the upper continuum. Obviously, there are also charged conjugated states localized near the negatively charged impurity, which rise from the lower continuum as $R$ increases. These states are occupied for sufficiently small $R$. Since $|v_n|=|v_p|$ at the point of maximal convergence $R=R_{m}$, the probability to find the electron near the negatively and positively charged impurities is then equal. As the distance between impurities $R$ increases further, the difference of the square moduli (\ref{difference}) changes sign because the Coulomb integral $C$ decreases with increasing $R$. This means that the electron wave function changes its localization to the positively charged impurity. This change of the wave function localization (the relocalization or "migration" effect \cite{footnote}) is explicitly shown in Fig.\ref{wf_LCAO} for $\zeta=0.85$ when $\epsilon_{0}=-0.453<0$. The value of $R_{m}$ in general depends on $r_0$ and on the value of the coupling constant $\zeta$. As the coupling increases, $R_{m}$ decreases (see Fig.\ref{crit_distance}).
It is clear that if $\epsilon_{0}>0$, then nothing interesting happens. Indeed, since $h_{pp}>0$ for $\epsilon_{0}>0$, we find that $|v_n|>|v_p|$ for any distance $R$ between the impurities. Therefore,
the wave function is always localized on the negatively charged impurity and, consequently, the wave function of the highest occupied state does not change its localization. Note that the behavior of the energy levels in this case given by green dash-dotted lines in Fig.\ref{fig2-LCAO} as
functions of $R$ is monotonous unlike the case where $\epsilon_0$ is negative.

The LCAO method used in this section has some limitations. For example, it cannot be applied for small distances $R$, and for values of charges when levels in the field of single centers enter continua. Therefore, we consider below the GK variational method which is free of these
limitations.

\section{Variational method} 
In this section, we use the GK variational method in order to find numerically solutions of the system of equations (\ref{system-equations}). The trial functions which belong to the class $\lambda=1$ (see the discussion above Eq.(\ref{system-equations})) and  satisfy the asymptotic at large distance are taken as
\begin{eqnarray}
\label{system2}
\phi(x,y)=A(x,y)\hspace{-1mm}\left(\sum\limits_{k=0}^{N}f_{2k}(x)y^{2k}+
i\hspace{-1mm}\sum\limits_{k=1}^{N^{'}}f_{2k-1}(x)y^{2k-1}\hspace{-1mm}\right)\hspace{-1mm},
\nonumber
\end{eqnarray}
\begin{eqnarray}
\hspace{-2mm}\chi(x,y)= A(x,y)\hspace{-1mm}\left(-i \sum\limits_{k=0}^{N}g_{2k}(x)y^{2k}
+\hspace{-1mm}\sum\limits_{k=1}^{N^{'}}g_{2k-1}(x)y^{2k-1}\hspace{-1mm}\right)\hspace{-1mm},
\label{ansatz}
\end{eqnarray}
where $A(x,y)=\exp[{-\sqrt{1-\epsilon^{2}}\sqrt{\left(|x|-R/2\right)^{2}+y^{2}+r_{0}^{2}}}]$.
These functions are substituted into Eq.(\ref{system-equations}) and their orthogonality to the residual with respect to the variable $y$ is required. Then we obtain the system of ordinary differential equations for the functions $f_k,g_k$.

The same method can be also used for solving the Dirac equation for the electron in the field of one Coulomb center with regularized potential. We found that approximate spectra determined by the GK method for the various number of terms in the ansatz $N,N^{'}$ describe the exact spectrum in the best way when $N=N'$.

\begin{figure}[ht]
  \centering
  \includegraphics[scale=0.32]{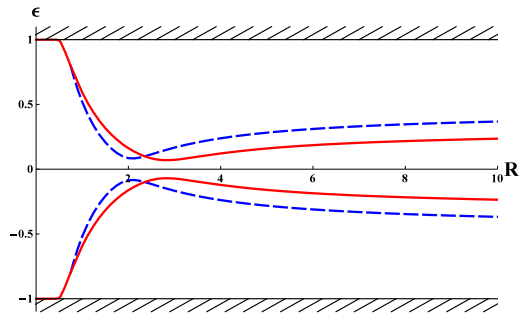}
  \caption{(Color online) The bound state energy levels in the electric dipole potential for $\zeta=0.85$
  and $r_0=0.05R_\Delta$ obtained in the LCAO (blue dashed lines) and variational (red solid lines)
  methods. }
  \label{fig2-variational}
\end{figure}

The approximate spectrum of the Dirac equation for the electron in the electric dipole potential is plotted in Fig. \ref{fig2-variational} for $\zeta=0.85$, $r_{0}=0.05 R_{\Delta}$ in the $N=1,N^{'}=1$ approximation (red solid line). For comparison, we plot in the same figure the energy levels obtained by using the LCAO technique (blue dashed line). Qualitatively both methods give similar results. For $\zeta=0.85$, the energy of the lowest electron bound state in the single positive Coulomb center problem (see, Fig.\ref{fig1-LCAO}) is negative (for the chosen charge $\zeta=0.85$ the single positive Coulomb center has only one negative energy level). Therefore, the wave function changes its localization on Coulomb impurities (see the video in the Supplemental Material \cite{suppl_mat}). Our analysis performed by making use of the GK variational
method confirms the conclusion made in the LCAO technique that the migration of the wave function of the highest energy occupied electron bound state takes place when the charges of impurities are such that the energies of the corresponding single Coulomb center problems cross.

The $N=1,N^{'}=1$ approximation in the GK method was also applied to the study of the spectrum
of an isolated impurity. A good agreement with the exact spectrum in Fig.\ref{fig1-LCAO} was obtained.
The value of the critical charge of the impurities (which form the electric dipole) depends on the regularization parameter. For $r_{0}=0.05R_{\Delta}$, the exact value of the critical charge is
$\zeta^{\rm exact}_{c}= 0.71$, and the approximate values determined by the GK method are $\zeta^{(00)}_{c}\approx 0.675$ and $\zeta^{(11)}_{c}\approx 0.68$ for the $N=0,N^{'}=0$ and $N=1,N^{'}=1$ approximations, respectively.

\section{Summary} 
In the present paper, we studied the Dirac equation for quasiparticles in gapped graphene with two oppositely charged impurities situated at a finite distance (finite electric dipole). By using the LCAO 
and variational GK methods, we found  the instability for quasiparticles in graphene in the supercritical electric dipole potential.  The found instability is connected with the change of localization of electron wave function on impurities and corresponds to the spontaneous creation of the electron-hole pair with the electron and hole screening the negatively and positively charged impurities, respectively. We showed that the necessary condition for the supercriticality to occur is the crossing of the energy levels of single positively and negatively charged impurities. Taken together these two levels should traverse the energy distance $2\Delta$, which is thus the necessary energetic condition for the supercriticality to occur.

The  instability of the supercritical electric dipole can be observed experimentally by placing  oppositely charged impurities with $\zeta>\zeta_c$ on graphene and then moving with a STM tip  one impurity toward the other one and afterwards moving the impurities apart again. The supercritical instability takes place if the impurities become partially screened that can be determined by measuring the local density of
states.

We are grateful to V.M. Loktev for useful discussions. This work is supported  by the Program of Fundamental Research of the Physics and Astronomy Division of the NAS of Ukraine.

\end{document}